\documentclass[conference]{IEEEtran}
\IEEEoverridecommandlockouts
\usepackage{cite}
\usepackage{amsmath,amssymb,amsfonts}
\usepackage[dvips]{graphicx}
\usepackage{textcomp}
\usepackage{xcolor}
\usepackage{comment}
\usepackage{commath}
\usepackage{algpseudocode}
\usepackage{steinmetz}
\usepackage{hyperref}
\usepackage{array}
\usepackage{booktabs}

\DeclareMathOperator*{\argmin}{\arg\!\min}
\DeclareMathOperator*{\argmax}{\arg\!\max}

\def\ninept{\def\baselinestretch{1.05}\let\normalsize\small\normalsize}
\ninept

\def\BibTeX{{\rm B\kern-.05em{\sc i\kern-.025em b}\kern-.08em
    T\kern-.1667em\lower.7ex\hbox{E}\kern-.125emX}}

\newcommand{\bh}{\boldsymbol{h}}
\newcommand{\by}{\boldsymbol{y}}
\newcommand{\bb}{\boldsymbol{b}}
\newcommand{\bn}{\boldsymbol{n}_p}
\newcommand{\bw}{\boldsymbol{w}}
\newcommand{\bH}{\boldsymbol{H}}
\newcommand{\bG}{\boldsymbol{G}}
\newcommand{\bI}{\boldsymbol{I}}
\newcommand{\bP}{\boldsymbol{P}_{\bH}}
\newcommand{\bPo}{\boldsymbol{P}_{\bH}^{\perp}}
\newcommand{\tr}{\mathrm{tr}}
\newcommand{\setC}{\mathbb{C}}
\newcommand{\setR}{\mathbb{R}}
\newcommand{\bphi}{\boldsymbol\varphi}
\newcommand{\brho}{\boldsymbol\rho}
\newcommand{\beps}{\boldsymbol{\epsilon}}
\newcommand{\bxi}{\boldsymbol{\xi}}
\newcommand{\btht}{\boldsymbol\theta}
\newcommand{\delphicp}{\nabla_{{\varphi}_{c,p}}}

\usepackage{soul}

\usepackage{etoolbox}
\makeatletter
\patchcmd{\@outputpage}{\if@figlist\@restonecoltrue\fi}{}{}{}
\makeatother

\usepackage{tikz}

\title{Estimating Multi-chirp Parameters using Curvature-guided Langevin Monte Carlo}
\author{Sattwik Basu, Debottam Dutta, Yu-Lin Wei, Romit Roy Choudhury\\ 
University of Illinois at Urbana-Champaign}

\begin{document}
\maketitle

\begin{abstract}
This paper considers the problem of estimating chirp parameters from a noisy mixture of chirps. 
While a rich body of work exists in this area, challenges remain when extending these techniques to chirps of higher order polynomials.
We formulate this as a non-convex optimization problem and propose a modified Langevin Monte Carlo (LMC) sampler that exploits the average {\em curvature} of the objective function to reliably find the minimizer.
Results show that our {\em Curvature-guided} LMC (CG-LMC) algorithm is robust and succeeds even in low SNR regimes, making it viable for practical applications.

\end{abstract}


\section{Introduction}
\label{sec:intro}

\begin{figure}[ht]
  \centering
  \vspace{-0.25in}
  \includegraphics[width=7cm, height=3.5cm]{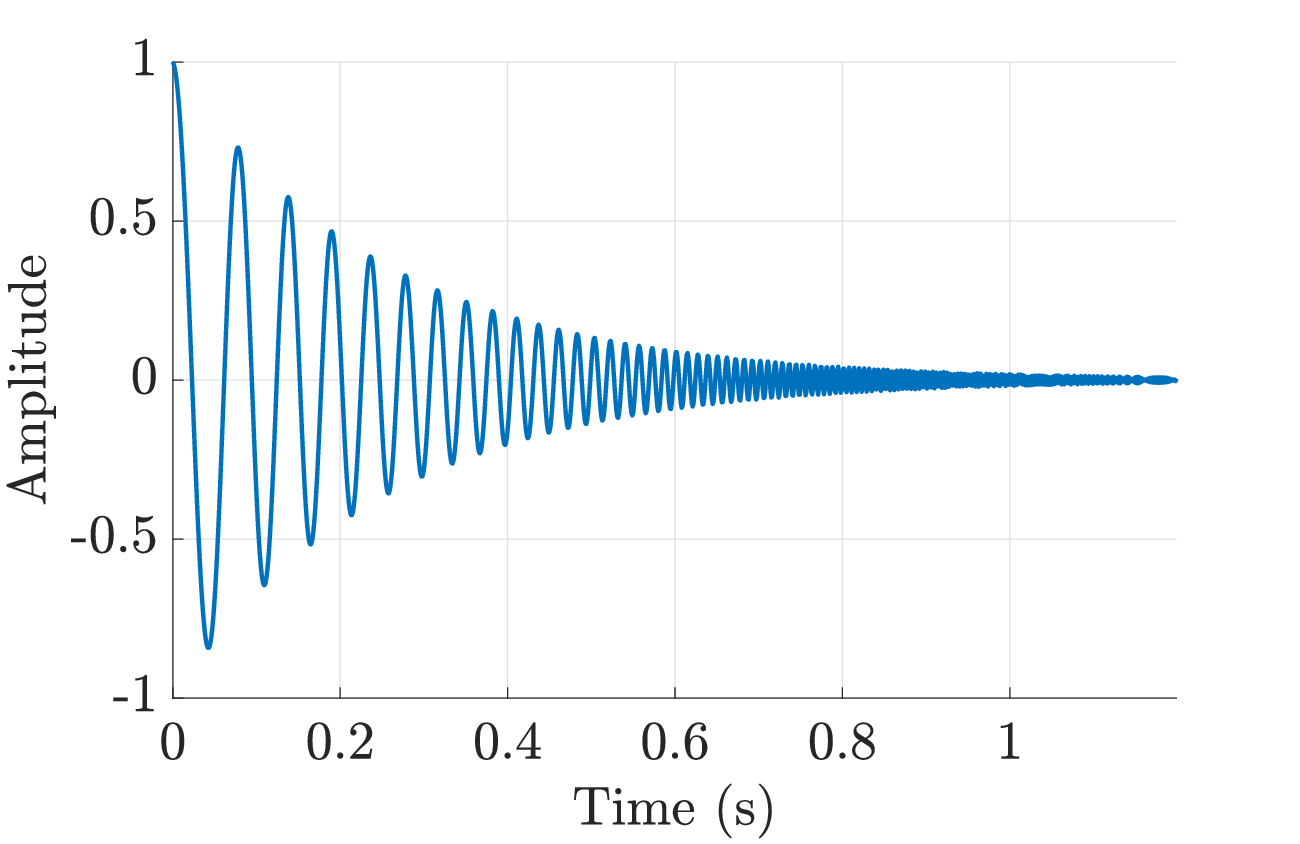}
  \vspace{-0.05in}
  \caption{Real part of a damped chirp with 4-th order polynomial IP.}
  \label{fig:basicChirp}
  \vspace{-1pt}
\end{figure}

Chirps are a fundamental class of non-stationary signals that underlie many applications. These signals are characterized by complex exponentials with a time-varying instantaneous phase (IP) and amplitude (IA), often modeled as quadratic or logarithmic functions. 
Of interest is the more complex scenario, where the IP and IA of a finite-duration chirp are $P^{\text{th}}$ and $A^{\text{th}}$ order polynomials, respectively. 
Equation \eqref{eqn:basicChirp} models such a chirp and Figure \ref{fig:basicChirp} visualizes it for $P = 4$ and $A = 3$.
Estimating parameters $\left\{\rho_a\right\}_{a=0}^{A}, \left\{\varphi_p\right\}_{p=1}^{P}$ of such signals becomes difficult when multiple chirps mix at low SNR regimes. This problem has been the focus of several studies in the past \cite{dpt,mchaf,phaf,djuric_kay, arun} and continues to be the subject of active research.
Our goal in this paper is to improve the state-of-the-art toward estimating mixtures of higher-dimensional chirps at lower SNRs.

\begin{align}\label{eqn:basicChirp}
    y(n) &= \underbrace{\sum_{a=0}^{A}\rho_{a}\left( \frac{n}{f_s}\right)^a}_\text{Instantaneous Amplitude (IA)} \hspace{-9pt} \exp\left\{j\smash[b]{ \underbrace{2\pi\sum_{p=1}^{P}\varphi_{p}\left( \frac{n}{f_s} \right)^p}_\text{Instantaneous Phase (IP)}} \right\}
\end{align}

Chirp parameter estimation is already fundamental to a wide range of applications such as wireless radars, audio processing, biomedical sensing, and astronomy. 
In radars, for instance, chirp mixtures are encountered when a transmitted chirp echoes back over a multipath channel from multiple moving targets. 
The IP continuously varies with time due to the relative motion between the reflectors and the receiver; estimating these parameters correctly reveals information about the position and movements of the targets. Another emerging application pertains to cardiac sounds, where the sound of blood flowing through heart valves has been shown to be a mixture of $2$ higher-order polynomial chirps \cite{pibarot, pibarot2}.
The vast majority of doctors find it difficult to manually separate these $2$ sounds when listening through a stethoscope \cite{steth}. 
Success in chirp parameter estimation could enable electronic stethoscopes that can isolate individual valve sounds in software.
This could obviously be valuable for downstream diagnosis and disease localization.

\textbf{Past work} on chirp parameter estimation has explored various lines of attack. 
Nonlinear transformations, like Higher-Order Ambiguity Functions (HAF, ML-HAFs, PHAFs) \cite{dpt, phaf}, iteratively transform chirps into sinusoids whose frequencies correspond to the desired chirp parameters. 
Despite their computational simplicity, HAFs struggle with identifiability issues in multi-component chirps with time-varying amplitudes, and show performance degradation in low-SNR conditions as errors in higher order parameters propagate down to the lower orders. Another body of work uses a maximum a posteriori (MAP)/maximum likelihood (MLE) framework to decouple the amplitude and phase parameters by solving a nonlinear least squares (NLS) problem \cite{friedlander_nss, fastmle, kundu2021chirp}. 
For example, \cite{sahakay} proposes a Monte Carlo approach to solve the NLS problem using importance sampling and a global optimization theorem \cite{pincus1968closed} closely linked with simulated tempering \cite{Marinari1992SimulatedTA} for chirp mixtures with $P =2$. The computational burden in this approach increases for higher order chirps because of the need to jointly sample from a discrete grid in parameter space. A more recent approach \cite{neri} takes the variational inference route to obtain an approximate posterior over the parameters of $P = 2$ chirp mixtures. 
A common limitation in most of these works is the performance degradation with increasing parameters ($P \times A \times N_c$, where $N_c$ is the number of chirps in the mixture). In our work, we focus on addressing these limitations.

\textbf{This paper} tackles mixtures of higher-order chirps ($P \geq 3$, $A \geq 2$, $N_c$ = $2$) where both IA and IP are polynomials.
We formulate this as a non-convex optimization problem whose objective function --- partially visualized in Figure \ref{fig:ObjFunc} --- contains a deep global minimum, numerous local minima clustered around the global minima, and relatively flat regions elsewhere.
We adopt a sampling approach to this optimization problem using a Langevin Monte Carlo (LMC) sampler.
While the stochastic behavior of LMC is designed to help in escaping local minima, we observed that the outcomes were unreliable in our case.
Past work in optimization \cite{jordan_lmc} has addressed this with a gradient smoothing approach which adds two benefits: (1) helps LMC overcome the local minima by smoothing the local landscape and (2) improves the gradient magnitude in far away regions so that the LMC sampler can progress towards the global minima, $\bphi_{opt}$.
Unfortunately, this (Gaussian) smoothing needs careful tuning; excessive smoothing can obscure the global minima, while inadequate smoothing retains the original problem.
In response to this, we propose a curvature-based smoothing method (utilizing the trace of the Hessian) to adaptively tune the Gaussian smoothing parameter throughout the whole optimization process.
The end result is that our {\em Curvature-guided} LMC optimizer (CG-LMC) is able to progress towards the global minima $\bphi_{opt}$ even from far-away regions, and as the optimizer comes close to $\bphi_{opt}$, it is able to tide over the surrounding stationary points to reliably converge near $\bphi_{opt}$.

\textbf{Experiments} with synthetic chirps at relatively low signal-to-noise (SNR) demonstrate a low estimation error and high reliability.
Our CG-LMC method outperforms two baselines, namely a classical Langevin Monte Carlo optimizer (LMC) \cite{rrt} and a Noise-Annealed Langevin Monte Carlo optimizer (NA-LMC) \cite{song}.

\section{Problem Formulation}
\label{sec:formulation}

\subsection{Multi-component Chirps}
\label{ssec: mc-chirps}

The signal model for an $N$-sample discrete-time multi-component chirp $y(n)$ sampled at a rate $f_s$ Hz is shown in Equation \eqref{eqn:mcchirp}.
\begin{align} \label{eqn:mcchirp}
    y(n) = \sum_{c = 1}^{N_c}A_c(n)\exp \left\{j2\pi \sum_{p=1}^{P}\varphi_{c,p}\left( \frac{n}{f_s} \right)^p \right\} + w(n)
\end{align}
where
\vspace{-0.1in}
\begin{align}
    A_c(n) &= \sum_{a=0}^{A_c}\rho_{c,a}\left( \frac{n}{f_s}\right)^a \quad \quad \text{and}\quad \quad
    w(n) \sim \mathcal{N}_\mathbb{C}(0, \sigma^2) \nonumber
\end{align}
for $n = 0, ... ,N-1$. Equation \eqref{eqn:mcchirp} represents a mixture of $N_c$ chirps polluted by Gaussian noise $w(n)$. Each chirp is assumed to have a $P$-th order phase polynomial and a $A_c$-th order amplitude polynomial. Estimating the parameters of the combined chirp entails estimating all the parameters shown in \eqref{eqn: paramvec}. 
\begin{align}\label{eqn: paramvec}
    \bphi = [\bphi_1 \, \dots \, \bphi_{N_c}]^T, \brho = [\brho_1 \, \dots \, \brho_{N_c}]^T 
\end{align}
Here, $\bphi_c = [\varphi_{c,1} \, \dots \, \varphi_{c,P}]$ and $\brho_c = [\rho_{c,0} \, \dots \, \rho_{c,A_c}]$ represent the phases and amplitudes of the $c$-th chirp, respectively. 

We can now rewrite the measured signal model in matrix-vector form to setup a nonlinear regression problem. 
\begin{align}
    \by = \bH(\bphi) \bb\left(\brho\right) + \bw
\end{align}

Here, $\by \in \setC^{N\times1}$ represents the measured signal, and the basis matrix, $\bH(\bphi) \in \mathbb{C}^{N\times \sum_{c=1}^{N_c} \left(A_c+1\right)}$ is comprised of the sub-matrices $\bH(\bphi_c) \in \setC^{N \times \left(Ac+1\right)}$ for $c = 1, ... , N_c$ such that
\begin{align}
\bH(\bphi) = \left[\bH(\bphi_1) \, \dots \, \bH(\bphi_{N_c})\right]
\end{align}

The columns of the $c$-th sub-matrix expressed compactly are $\bH(\bphi_c) = \left[\bh_{c,0} \, \dots \, \bh_{c,A_c}\right]$ where $\bh_{c,a} \in \setC^{N \times 1}$ is
\begin{align}
\bh_{c,a} =
    \begin{bmatrix}
    \left(\frac{0}{f_s}\right)^a \exp \left\{j2\pi \sum_{p=1}^{P}\varphi_{c,p} \left(\frac{0}{f_s} \right)^p \right\}\\[2ex]
    \vdots \\[2ex]
    \left(\frac{N-1}{f_s}\right)^a \exp \left\{j 2\pi \sum_{p=1}^{P}\varphi_{c,p}\left(\frac{N-1}{f_s} \right)^p \right\}\\[2ex]
    \end{bmatrix}
\end{align}
for $a = 0, \dots, A_c$. The vector $\bb\left(\brho \right) \in \setC^{\sum_{c=1}^{N_c} \left(A_c+1\right) \times 1}$ consists of amplitude parameters such that 
\begin{align}
\bb\left(\brho \right) = \left[\bb \left(\brho_{1}\right) \, \dots \, \bb\left(\brho_{N_c}\right) \right]^T
\end{align}
with $\bb \left(\brho_{c}\right) = \left[\rho_{c,0} \, \dots \, \rho_{c,A_c} \right]$ for $c = 1, \dots , N_c$. 
From here on, we denote $\bH(\boldsymbol\varphi)$ and $\bb\left(\boldsymbol{\rho}\right)$ as $\bH$ and $\bb$ respectively, for the sake of brevity. 

\subsection{MAP estimation}
\label{ssec: MLE}

Since $\bw$ is assumed to be white Gaussian noise, the likelihood function can be expressed as
\begin{align}
p(\boldsymbol{y}; \boldsymbol\varphi_, \boldsymbol{b}) = \frac{1}{(2\pi)^{N/2}\sigma^{N}}\exp \left(- \frac{1}{2\sigma^2}\norm{\by - \bH \bb}^2_2 \right)
\end{align}

\begin{figure}[t]
\vspace{-0.04cm}
    \centering
          \includegraphics[scale=0.32]{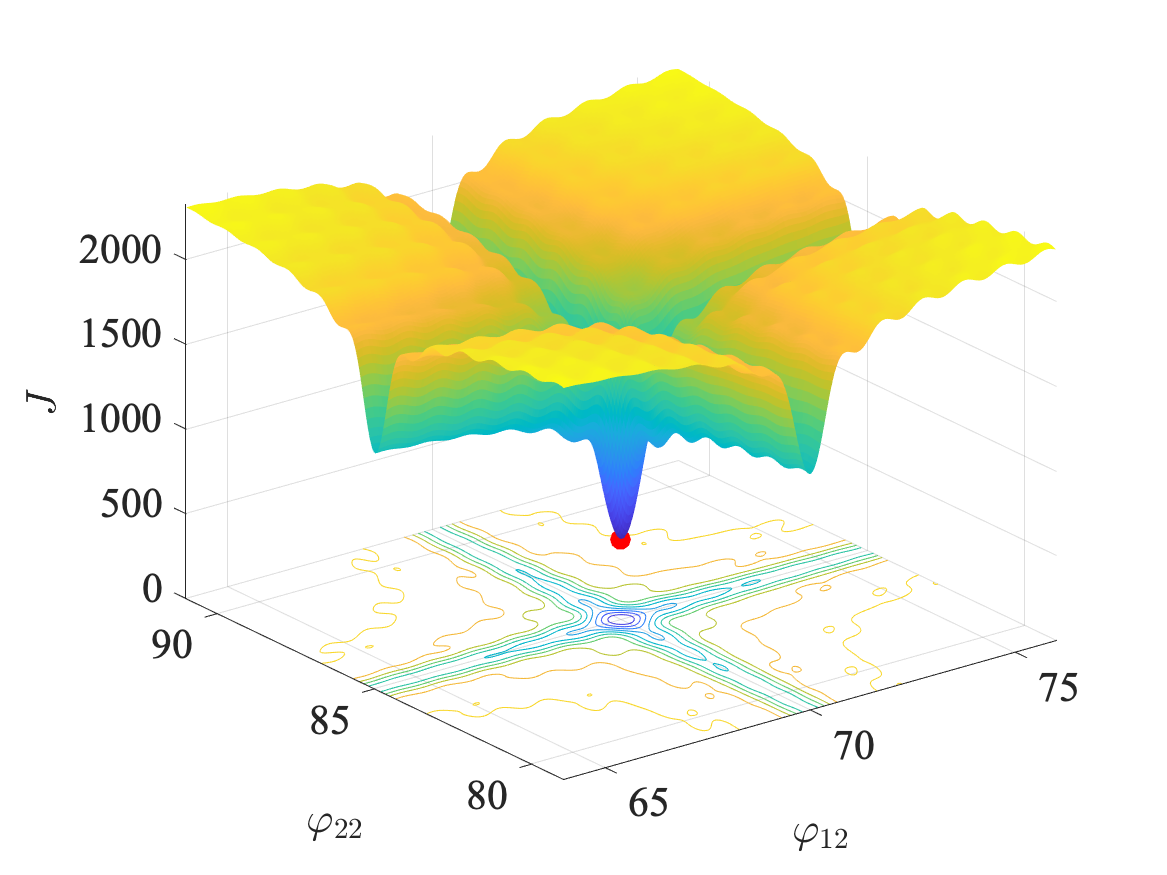}
          \vspace{-5pt}
    \caption{A slice of the non-convex 4-D objective function for a two-chirp mixture with $P=2$. Parameters $\varphi_{12}$, $\varphi_{22}$ are varied while others are fixed at optimum values.}
    \label{fig:ObjFunc}
\vspace{-10pt}
\end{figure}

To obtain MAP estimates of vectors $\boldsymbol\varphi$ and $\boldsymbol{\rho}$ with uniform priors, we see that
\begin{align}
 \argmax_{\bphi, \brho} \ln p\left(\bphi, \brho | \by\right) \label{posterior} &= \argmax_{\bphi, \brho} \ln p\left(\by|\bphi, \brho\right) + \ln p\left(\bphi, \brho \right) \\
 &= \argmin_{\bphi, \brho} \norm{\by - \bH \bb}^2_2 \label{eqn: objFunc}
\end{align}

\subsection{Nonlinear Least Squares}
\label{ssec: NLS}

Minimizing Equation \eqref{eqn: objFunc} is complicated due to the unknown nonlinear dependency on $\bphi$. However, noting that the NLS criterion is quadratic in $\bb$, for any given vector $\bphi$, $\hat{\bb} =  \bG^{-1} \bH^*\by$ where, $\bG = \left( \bH^*\bH + \gamma \bI\right)$. Here, $\gamma$ is a regularization parameter and $^*$ represents the Hermitian transpose of a complex matrix. 

Substituting $\hat{\bb}$ back into Equation \eqref{eqn: objFunc} yields a modified cost function $J: \setR^{N_cP} \rightarrow \setR$ that depends only on $\boldsymbol\varphi$.
\begin{align}
J(\boldsymbol\varphi) &= \norm{\by - \bH \hat{\bb}}^2_2 = \by^* \bPo \by
\end{align}

Minimizing this objective function gives us the optimal phase vector $\bphi_{opt}$. That is,
\begin{align}
\bphi_{opt} &=  \argmin_{\bphi} J(\bphi) = \argmin_{\bphi} \boldsymbol{y}^*\bPo \by \label{eqn: objfunc}
\end{align}
Here, $\bPo = \bI - \bP$ with $\bP = \bH \bG^{-1} \bH^*$. Once the optimal nonlinear parameters $\bphi_{opt}$ are estimated, $\brho$ can be computed via least squares. However, optimizing $J(\bphi)$ in Equation \eqref{eqn: objfunc} is not trivial as it is highly non-convex (Fig. \ref{fig:ObjFunc}). 
The complexity grows 
with increasing dimensions of the parameter space.

\section{Curvature-Guided Langevin Monte Carlo (CG-LMC)}
\label{sec:sampling2optimize}

\subsection{Background}
\label{ssec: Background}
Recent works have demonstrated that optimization problems can be addressed by sampling from a distribution whose mode is concentrated around the global optimum. Monte Carlo algorithms such as importance sampling and Metropolis-Hastings \cite{robertcasella_mcmc, luengo_mcmc} are well-known for this purpose. However these algorithms are problematic in high-dimensional settings due to high rejection rates and difficulties in designing proposal distributions \cite{speagle_mcmc}. Moreover, the distribution associated with the non-convex objective functions are multi-modal adding further complexity to the optimization. In these scenarios, Langevin Monte Carlo (LMC) offers attractive benefits \cite{neal_hmc_book}. 

LMC has found increasing interest in the nonconvex optimization literature  \cite{rrt, samplingFasterOpt, lmc_global_conv, dalalyan_opt_lmc} due to its resemblance with Stochastic Gradient Descent (SGD) \cite{sgd_ml, welling_lmc_sgd, pml_murphy}. The core idea is to simulate a continuous-time Langevin Diffusion \cite{diffusionbook} as shown in Equation \eqref{eqn:langevinDiff}. 
\begin{align} \label{eqn:langevinDiff}
\text{d}\boldsymbol{X}\left(t\right) = -\nabla F \left(\boldsymbol{X}\left(t\right)\right)\text{d}t + \sqrt{2\beta^{-1}} \text{d}\boldsymbol{B}(t), t \geq 0
\end{align}
where the distribution of $\boldsymbol{X}\left(t\right) \in \setR^d$ as $t \rightarrow \infty$ converges to a stationary distribution $\pi\left( . \right) \propto \exp \left(-\beta F \left( . \right) \right)$ \cite{lmcstatdist}. 
Here, $F$ is called the potential function, $\{\boldsymbol{B}(t)\}_{t \geq 0}$ is a Brownian process in $\setR^d$, and $\beta > 0$ is an inverse temperature parameter. For sufficiently large $\beta$, drawing samples from $\pi(.)$ produces values close to the minimizer of $F$ with high probability.

For simulating Equation \eqref{eqn:langevinDiff} the Euler-Maruyama \cite{kloeden_sde} scheme is used to obtain the discrete-time Markov chain (LMC)
\begin{align} \label{eqn:lmc}
\boldsymbol{X}^{(k+\eta)} &= \boldsymbol{X}^{(k)} - \eta \nabla F\left(\boldsymbol{X}^{(k)} \right) + \sqrt{2\eta\beta^{-1}} \boldsymbol{\xi}^{(k)}
\end{align} 
where $\eta > 0$ is the stepsize, $k$ the iteration index and $\boldsymbol{\xi}^{(k)} \sim \mathcal{N}(0, \boldsymbol{I}_d) $. Following each step, the Metropolis-Hastings algorithm is used to either accept or reject the proposed sample $\boldsymbol{X}^{(k+\eta)}$ \cite{mala}. The interplay between gradient and the stochastic perturbation helps in overcoming local stationary points to improve convergence. As $k$ increases, the iterates $\boldsymbol{X}^{(k+\eta)}$ converge on the minimizer of $F$.

\subsection{LMC for optimizing $J(\boldsymbol\varphi)$}
\label{ssec: LMC}
To apply LMC to optimize $J(\boldsymbol\varphi)$ in Equation \eqref{eqn: objfunc}, we define $\pi(\bphi) \propto \exp\left( -\beta J(\bphi) \right)$ as the stationary distribution. Observe that $\pi(\bphi)$ is the marginalized posterior $p(\bphi | \by)$ in Equation \eqref{posterior} representing the decoupling of $\bphi$ from $\brho$. Since $J(\bphi)$ is non-convex, $\pi(\bphi)$ is multi-modal with the largest mode around $\bphi_{opt}$. To draw samples from this distribution, we use the Markov chain
\begin{align} \label{eqn:lmcPhi}
\bphi^{(k+1)} = \bphi^{(k)} - \eta \nabla_{\bphi} J\left(\bphi^{(k)} \right) + \sqrt{2\eta\beta^{-1}} \boldsymbol{\xi}^{(k)}
\end{align}
where $\boldsymbol{\xi}^{(k)} \sim \mathcal{N}(0, \boldsymbol{I}_{N_cP})$ and $\bphi^{(0)} \sim \pi^{(0)}$, an initial distribution. The gradient elements of $\nabla_{\bphi} J \in \setR^{N_cP \times 1}$ (derivation shown in the appendix) contains partial derivatives $\nabla_{{\varphi}_{c,p}} J$ where
\begin{align} \label{eqn:grad}
\nabla_{{\varphi}_{c,p}} J = -2\mathbf{Re}\left[\by^* \bPo \nabla_{{\varphi}_{c,p}}\bH \bG^{-1}\bH^*\by\right] 
\end{align}
for $c = 1 \, \dots \, N_c$ and $p = 1 \, \dots \, P$. The term $\nabla_{{\varphi}_{c,p}}\bH$ in Equation \eqref{eqn:grad} contains element-wise partial derivatives of $\bH$ with respect to each parameter $\varphi_{c,p}$ for $c = 1 \, \dots \, N_c$ and $p = 1 \, \dots \, P$. 
\begin{align}
    \delphicp\bH(\bphi) &= \left[\delphicp\bH(\bphi_1) \, \dots \, \delphicp\bH(\bphi_{N_c})\right] \nonumber \\
    &= \left[\boldsymbol{0} \, \dots \delphicp \bH(\bphi_c) \dots \, \boldsymbol{0} \right] \nonumber \\
    &= \left[\boldsymbol{0} \, \dots \, j 2\pi \bn \odot \left[\bh_{c,0} \, \dots \, \bh_{c,A_c}\right]  \dots \boldsymbol{0} \right]
\end{align}
where $\bn = \left[\left(\tfrac{0}{f_s}\right)^p \, \dots \, \left(\tfrac{N-1}{f_s}\right)^p \right]^T$ and $\odot$ denotes the element-wise product.

Although LMC offers promise, it is not always able to ``jump out'' of a local minima, despite it's stochastic nature \cite{simTemper}.
Such issues become pronounced in higher dimensional chirps where the objective function has many complex ripples.
Smoothing techniques have been proposed \cite{song} to help the LMC sampler combat such cases.

\begin{table*}[htbp]
\centering
\caption{Comparison between CG-LMC, LMC, and NA-LMC showing the mean $\pm$ SD of the estimated parameter for different SNRs.}
\vspace{-3pt}
\label{3dB_mean_std_abs}
\scriptsize 
\begin{tabular}{|c|p{2cm}|p{2cm}|p{2cm}|p{2cm}|p{2cm}|p{2cm}|}
\hline
\textbf{Algorithm} & \multicolumn{2}{|c|}{\textbf{CG-LMC}} & \multicolumn{2}{|c|}{\textbf{LMC}} & \multicolumn{2}{|c|}{\textbf{NA-LMC}} \\
\hline
\textbf{Parameters} & \textbf{\centering 3 dB} & \textbf{\centering 12 dB} & \textbf{\centering 3 dB} & \textbf{\centering 12 dB} & \textbf{\centering 3 dB} & \textbf{\centering 12 dB} \\
\hline
$\varphi_{1,1} = 10$ & 10.17 ± 0.55 & 9.89 $\pm$ 0.59 & 10.05 ± 0.97 & 9.69 $\pm$ 2.11 & 9.93 ± 1.58 & 9.67 $\pm$ 0.81 \\
\hline
$\varphi_{1,2} = 40$ & 40.85 ± 1.88 & 44.91 $\pm$ 0.86 & 39.98 ± 8.71 & 43.67 $\pm$ 13.59 & 41.20 ± 10.93 & 41.58 $\pm$ 6.58 \\
\hline
$\varphi_{1,3} = -70$ & -68.96 ± 3.91 & -76.32 $\pm$ 1.81 & -80.43 ± 24.51 & -83.22 $\pm$ 5.07 & -77.04 ± 5.63 & -77.44 $\pm$ 2.74 \\
\hline
$\varphi_{1,4} = 110$ & 109.84 ± 2.40 & 104.66 $\pm$ 5.34 & 134.39 ± 21.57 & 125.64 $\pm$ 10.21 & 126.36 ± 24.37 & 118.78 $\pm$ 23.59 \\
\hline
$\varphi_{2,1} = 50$ & 50.08 ± 0.55 & 49.98 $\pm$ 0.21 & 50.55 ± 0.92 & 50.21 $\pm$ 0.91 & 50.27 ± 1.20 & 51.08 $\pm$ 1.18 \\
\hline
$\varphi_{2,2} = 60$ & 62.58 ± 1.27 & 60.68 $\pm$ 2.17 & 60.33 ± 4.74 & 61.85 $\pm$ 4.91 & 57.34 ± 9.02 & 57.77 $\pm$ 9.89 \\
\hline
$\varphi_{2,3} = -90$ & -104.46 ± 7.05 & -93.23 $\pm$ 1.28& -102.16 ± 7.19 & -106.05 $\pm$ 16.29 & -85.59 ± 18.26 & -79.32 $\pm$ 22.61 \\
\hline
$\varphi_{2,4} = 105$ & 112.26 ± 3.91 & 107.11 $\pm$ 12.06& 129.99 ± 17.15 & 130 $\pm$ 26.03 & 106.80 ± 8.20 & 119 $\pm$ 21.02 \\
\hline
\end{tabular}
\vspace{-10pt}
\end{table*}

\subsection{Gaussian-smoothed Langevin Monte Carlo}
\label{ssec: GS-LMC}
Given the landscape of our objective function $J(\boldsymbol\varphi)$, the initial points located far from the global minimum have negligible gradients while those initialized nearby risk being trapped in local stationary points. 
An effective solution is to add additional noise during Langevin updates to perturb the gradients \cite{song}. Adding noise has an implicit gradient smoothing effect as explained in \cite{nesterov2017random, jordan_lmc}. To see this, we write a modified LMC as
\begin{align} \label{eqn:smoothing}
\bphi^{(k+1)} &= \bphi^{(k)} - \eta \nabla_{\bphi} J\left(\bphi^{(k)} \right) + \sqrt{2\eta\beta^{-1}} \boldsymbol{\xi}^{(k)} + \sigma \beps^{(k)}
\end{align}
where $\beps^{(k)} \sim \tilde{\pi}\left(\beps^{(k)}\right) = \mathcal{N}(0, \boldsymbol{I}_{N_cP})$ is another Gaussian perturbation vector independent from $\bxi^{(k)}$ with variance adjusted by $\sigma$. Now, by setting $\btht^{(k+1)} = \bphi^{(k+1)} - \sigma \beps^{(k)}$ followed by taking the expectation with respect to $\beps^{(k-1)}$, we get
\begin{align} 
\mathbb{E}_{\beps^{(k-1)}}\left[\btht^{(k+1)}\right] &= \btht^{(k)} - \eta \mathbb{E}_{\beps^{(k-1)}}\left[ \nabla_{\bphi} J\left(\btht^{(k)} + \sigma \beps^{(k-1)} \right)\right] \nonumber \\
   & \qquad \qquad + \sqrt{2\eta\beta^{-1}} \bxi^{(k)} 
\label{eqn:gausssmooth}
\end{align}
The expectation term in the R.H.S of Equation \eqref{eqn:gausssmooth} is a convolution between $\nabla_{\bphi} J\left(\bphi_{k} \right)$ and a Gaussian pdf. This is referred to as the Gaussian-smoothed gradient $\nabla_{\bphi} J_\sigma$ \cite{nesterov2017random}. 
As a result of smoothing, the iterates $\btht^{(k+1)}$, on average, move along a smoothed gradient field minimizing the traps due to local minima and saddle points.

\subsection{Need for Curvature-guided Gaussian-smoothing}
\label{ssec: curvature}

It is crucial to note that the choice of $\sigma$ determines the extent of smoothing, which in turn guides convergence. 
To show this, we focus on the expectation term in Equation \eqref{eqn:gausssmooth}. 
\begin{align}
\nabla_{\bphi} J_\sigma &= \int \nabla_{\bphi} J\left(\btht^{(k)} + \sigma \beps^{(k-1)} \right) \tilde{\pi}\left(\beps^{(k-1)}\right) d\beps^{(k-1)} \\ 
&= \frac{1}{\sigma} \int \nabla_{\bphi} J\left(\bphi^{(k)}\right) \tilde{\pi}\left( \frac{\bphi^{(k)} - \btht^{(k)}}{\sigma}\right) d\bphi^{(k)} \label{eqn:ip}
\end{align}
Equation \ref{eqn:ip} represents an inner product between the gradient $\nabla_{\bphi} J$ and a kernel $\tilde{\pi}$ that is scaled by $\sigma$ and translated by $\btht^{(k)}$. 
A large $\sigma$ widens the kernel $\tilde{\pi}$, smoothing over wide regions of the parameter space, which can obscure the global minimum. A smaller $\sigma$ sharpens the kernel smoothing only over a small local neighborhood.
Thus, for reliable convergence, $\sigma$ needs to be adaptive; large when far away from the global minima so that the flat landscape can tilt towards the global minima (see Fig. \ref{fig:ObjFunc}).
When arriving closer to the global minima, $\sigma$ needs to reduce so that only the local minima is smoothened without destroying the natural curvature of the landscape. 

Without knowledge of the global minima, adapting $\sigma$ is difficult. 
Past work \cite{song} have empirically scheduled $\sigma$ to start as a large value and decrease in predefined steps as iterations progress.

We propose to leverage the average curvature of the objective function to regulate this adaptation. 
We obtain curvature information from $\tr\left\{\nabla_{\bphi}^2 J\right\}$, i.e., the trace of the Hessian of $J$, and update $\sigma$ as:  
\begin{align} \label{eqn:sigmaupdate}
\sigma^{(k+1)} = \max\left[\sigma_{\min}, \sigma^{(k)} - \mu_\sigma \left| \tr\left\{\nabla_{\bphi}^2 J(\bphi^{(k)})\right\} \right| \right]
\end{align}
where $\mu_\sigma$ is the step size, $k$ is the iteration index and $\sigma_{min}$ is a user defined minimum value.
The Hessian, $\nabla_{\bphi}^2 J$ in \eqref{eqn:sigmaupdate} is approximated using Stein's Lemma \cite{bbohess} with the same perturbation vector $\beps^{(k)}$ that is used to compute $\nabla_{\bphi} J_\sigma$. 
In addition, since we require the magnitude of average curvature and not the concavity of $J$, we use the absolute value of the trace. 
We noticed that a recent work \cite{slgh} has used the trace as well but in the context of Gaussian Homotopy.

\textbf{Initialization.}
Our algorithm begins with multiple randomized, diffused initializations, priming the LMC algorithm using shorter-length signals before transitioning to the full-length measured signal $\by$. 
These shorter signals correspond to objective functions with broader global minima \cite{angeby_nils}, providing a more reliable starting point for the lower order phase parameters. This approach eliminates dependency on additional initialization algorithms used in prior works.

\section{Experiments and Results}
\label{sec:eval}

We compare \textbf{CG-LMC} against two baselines, \textbf{LMC} and \textbf{NA-LMC}.
We test $5$ different chirp mixtures ($N_c=2$) with all chirps parameterized by $A=3$ and $P=4$. The chirp signals are $1.0$s long, sampled at $f_s$ = $1000$ Hz, and their $\bphi$ and $\brho$ values are selected arbitrarily while ensuring that the instantaneous frequencies never exceed $f_s/2$. 
Gaussian noise is added to achieve $3$ dB and $12$ dB SNRs. 
We report the accuracy of the estimated chirp parameters i.e., mean and standard deviation (SD) over $5$ runs; we also show the convergence behavior by tracking different initializations near and far from the global minima. 
Our MATLAB code is available at \url{https://github.com/basusattwik/ChirpEstimation}

\textbf{Table \ref{3dB_mean_std_abs}} reports the phase parameter estimates for one specific chirp with detailed comparisons.
At $3$ dB SNR, the mean of the phase parameter estimates from CG-LMC are closer the true value with smaller standard deviations, than those of LMC or NA-LMC, except in parameters $\varphi_{2,3}, \varphi_{2,4}$. 
This demonstrates the benefit of curvature-guided Gaussian smoothing as the low SNR regimes have more jagged objective functions; LMC in contrast tends to exhibit higher error and NA-LMC tends to exhibit higher SD.
At a higher SNR of $12$ dB, CG-LMC shows improved estimates with smaller variance compared to both LMC and NA-LMC.

\begin{figure}[ht]
    \centering
          \vspace{-8pt}
          \includegraphics[scale=0.35]{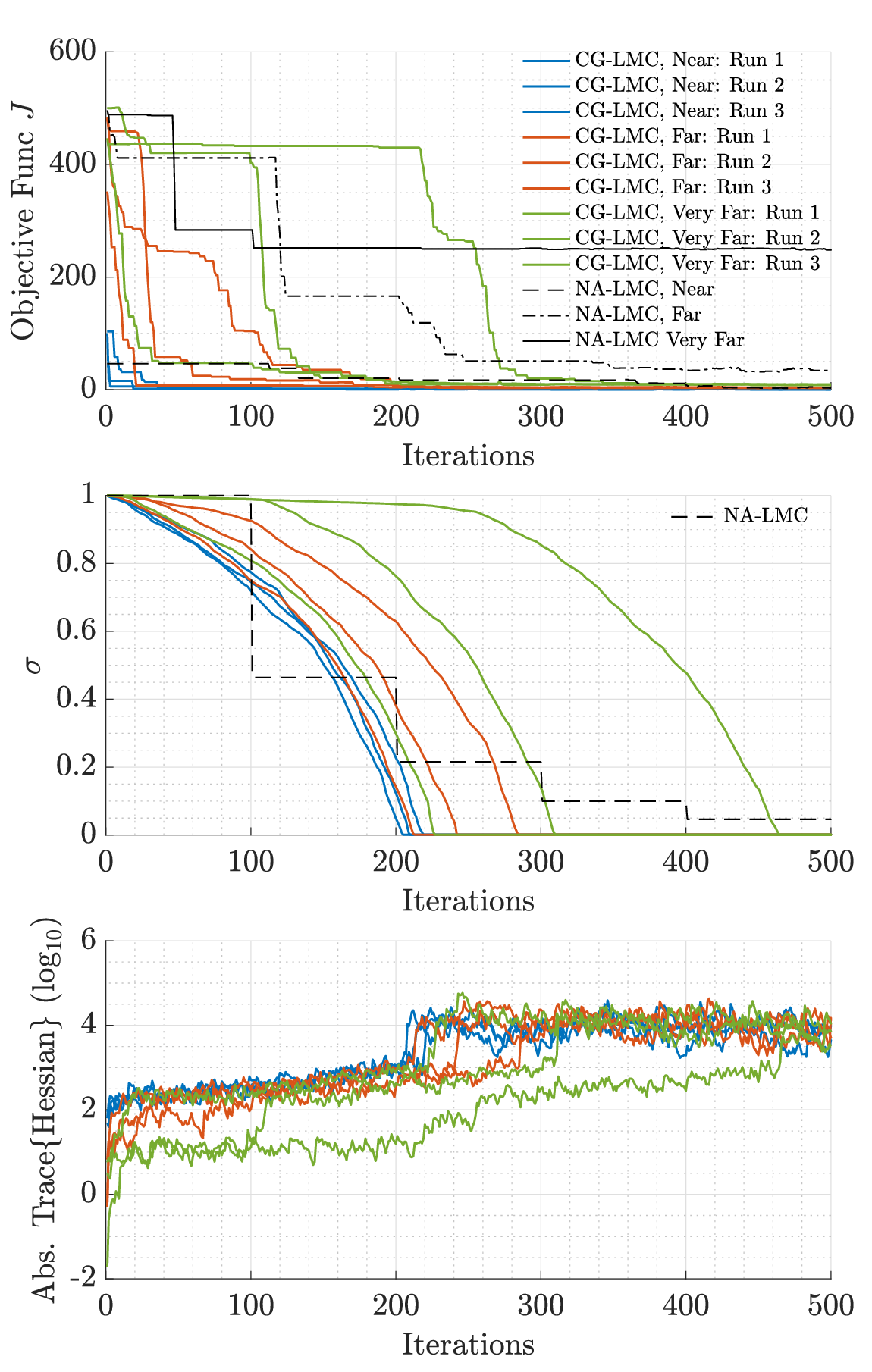}
          \vspace{-8pt}
    \caption{(a) 9 sample runs in CG-LMC and 3 in NA-LMC, (b) Comparing $\sigma$ adaptation in CG-LMC vs. the fixed schedule in NA-LMC, (c) Visualizing $\tr\left\{\nabla_{\bphi}^2 J\right\}$ over different CG-LMC runs.}
    \label{fig:InnerWorking}
\vspace{-17pt}
\end{figure}

\textbf{Fig. \ref{fig:InnerWorking}} sheds light on the internal behavior of CG-LMC for the same experiment as above.
We track a total of $9$ CG-LMC sample runs with the first $3$ initialized close to the global minimum (blue curves), $3$ moderately far (orange curves), and $3$ very far (green curves) from the global minimum.
For comparison, $3$ sample runs of NA-LMC are similarly initialized, near, far and very far away. 
Fig. \ref{fig:InnerWorking}(a) shows how CG-LMC is able to reliably converge to the global minima regardless of their initialization.
Of course, far away initializations tend to take longer compared to nearby ones.
In contrast, NA-LMC struggles to converge when initialized far away, leading to high SD.

\textbf{Fig. \ref{fig:InnerWorking}(b) and (c)} further show the smoothing parameter $\sigma$ and $\tr\left\{\nabla_{\bphi}^2 J\right\}$ for CG-LMC and NA-LMC.
Since NA-LMC is not adaptive, $\sigma$ reduces over iterations as a step function (hence, a single line in Figure \ref{fig:InnerWorking}(b)). 
While CG-LMC and NA-LMC both reduce $\sigma$, observe that CG-LMC waits far longer.
This allows CG-LMC to smoothen over large areas when the iterates are far from the global minima, helping them move in the correct direction.
This can be validated from Fig. \ref{fig:InnerWorking}(c), where the curvature is low for the green curves over the initial iterations.
Once the curvature begins to increase, the value of $\sigma$ begins to fall.
On the other hand, the sample runs that were initialized near and moderately far (blue and red), show a rapid decrease in $\sigma$ owing to the already large curvature near the global minimum. 
In contrast, since NA-LMC is ``blind'' to the landscape of the objective function, the method of step-wise smoothing may or may not guide the optimization to the global minima.
This negatively impacts the reliability (or SD) of estimation.

\textbf{Additional chirp mixtures} were designed at 3 dB SNR using different $\bphi$ and $\brho$ values. 
The mean absolute errors (ME) in the estimated phase parameters were [1.03, 3.55, 5.12, 11.24, 3.59, 4.19, 7.75, 7.87] over 5 runs while NA-LMC showed [0.77, 5.73, 13.33, 24.30, 4.67, 23.50, 12.26, 14.06].   
For CG-LMC, errors are close to those shown in Table \ref{3dB_mean_std_abs} demonstrating a generalization of performance across a range of parameters. 
In addition, the low values of ME in CG-LMC indicates robust estimation at low SNRs.

\section{Conclusions and Future Work}
\label{sec:conclusion}
We propose CG-LMC, an optimization algorithm for estimating multi-chirp parameters. We build on existing methods to optimize the chirp-centric objective function and show that leveraging the intrinsic curvature of the function can be valuable, especially in making the algorithm relatively robust to initialization. Future work includes comparing the estimators with the Cram{\'e}r-Rao bounds and exploring source separation methods to isolate chirps for parameter estimation in a lower-dimensional space. 
We leave these to future research.

\section{Acknowledgements}
\label{sec:conclusion}
We thank Foxconn and NSF (grant 2008338, 1909568, 2148583, and MRI-2018966) for funding this research. 
We are also grateful to the reviewers for their insightful feedback.

\clearpage

\newpage

\bibliographystyle{ieeetr}
\bibliography{conference_main}

\clearpage

\section*{Appendix}
\subsection{Gradient of the Objective Function}
\label{ssec: Grads}
We derive a closed form expression of $\nabla_{\boldsymbol{\varphi}} J \in \setR^{N_cP \times 1}$, the gradient of $J$ with respect to the parameters $\bphi$. We employ matrix differentials for conveniently obtaining the gradients without involving the complexity of tensors. Starting with the differential of the objective function in Equation \eqref{eqn: objfunc}, we see that
\begin{align}
dJ &= d\left(\by^* \bPo \by\right) \nonumber \\
   &= d\by^* \bPo \by + \by^* d\bPo \by + \by^* \bPo d\by \nonumber \\
   &= \by^* d\bPo \by \nonumber \\
   &= \tr \{d\bPo \by \by^*\} \label{eqn:dJ'_1}
\end{align} 
Next, we expand $d\bPo$ using standard expressions for the differentials of matrix products and inverses. This gives us,
\begin{align}
    d\bPo &= d\left(\bI - \boldsymbol{P}_{\bH} \right) \nonumber \\
          &= -d\boldsymbol{P}_{\bH} \nonumber \\
          &= -d\left( \bH \bG^{-1} \bH^* \right) \nonumber \\
          &= -d\bH \bG^{-1} \bH^*  - \bH d\bG^{-1} \bH^* \nonumber \\
    & \qquad \qquad - \bH \bG^{-1} d\bH^* \nonumber \\
          &= -d\bH \bG^{-1} \bH^*  \nonumber \\
    & \qquad \left. - \bH \left\{-\bG^{-1} \left(d\bG\right) \bG^{-1} \right\} \bH^* \right. \nonumber \\
    & \qquad \qquad - \bH \bG^{-1} d\bH^* \nonumber \\
           &= - d\bH \bG ^{-1} \bH^* \nonumber \\
    & \qquad + \bH \bG^{-1} \left(d\bH^*\bH + \bH^* d\bH \right) \bG^{-1} \bH^* \nonumber \\
    & \qquad \qquad - \bH \bG^{-1} d\bH^* \nonumber \\
           &= - (\bI - \bP) d\bH \bG^{-1}\bH^* \nonumber \\
    & \qquad \qquad - \bH \bG^{-1} d\bH^* (\bI - \bP) \nonumber \\
           &= - \bPo d\bH \bG^{-1}\bH^* \nonumber \\
    & \qquad \qquad - \bH \bG^{-1} d\bH^* \bPo \label{eqn:dPH_perp} 
\end{align}
The expression in Equation \eqref{eqn:dPH_perp} can be substituted back into Equation \eqref{eqn:dJ'_1} to obtain the full differential $dJ$. We then apply the cyclic property of the matrix trace to further simplify this expression and move towards finding an expression of the gradient.
\begin{align}
    dJ &= \tr \left[\left(-\bPo d\bH \bG^{-1}\bH^* \right. \right. \nonumber \\
    & \qquad \qquad \left. \left. - \bH \bG^{-1} d\bH^* \bPo \right) \by \by^* \right] \nonumber \\
    &= -\tr \left[\bPo d\bH \bG^{-1}\bH^* \by \by^*\right] \nonumber \\
    & \qquad \qquad - \tr \left[ \left( \bPo d\bH \bG^{-1}\bH^*\right)^* 
    \by \by^*\right] \nonumber \\
    &= -\tr \left[\bPo d\bH \bG^{-1}\bH^* \by \by^*\right] \nonumber \\
    & \qquad \qquad - \left\{\tr \left[\by \by^* \bPo d\bH \bG^{-1}\bH^* \right]\right\}^* 
    \nonumber \\
    &= -\by^* \bPo d\bH \bG^{-1}\bH^*\by \nonumber \\
    & \qquad \qquad - \left\{\by^* \bPo d\bH \bG^{-1}\bH^*\by \right\}^* \nonumber \\
    &= -2\mathbf{Re} \left[\by^* \bPo d\bH \bG^{-1}\bH^*\by \right] \label{eqn:dJ'_2}
\end{align}
where $\mathbf{Re}(.)$ denotes the real part of a complex number. 

To further simplify, we expand the differential $d\bH$ in Equation \eqref{eqn:dJ'_2} on the basis of its gradient through a first order approximation, i.e., $d\bH = \sum_{c = 1}^{N_c}\sum_{p=1}^{P} \nabla_{{\varphi}_{c,p}} \bH d\varphi_{c,p}$. We use $\delphicp(.)$ to denote the partial derivative with respect to the parameter $\varphi_{c,p}$. Since the differential $d\varphi_{c,p}$ is simply a scalar, it can be moved to the end of the double-sum. By rearranging the terms, we get 
\begin{align}
dJ &= -2\mathbf{Re} \left[\by^* \bPo \left(\sum_{c = 1}^{N_c}\sum_{p=1}^{P} \nabla_{{\varphi}_{c,p}} \bH d\varphi_{c,p} \right) \bG^{-1}\bH^*\by \right] \nonumber \\
&= \sum_{c = 1}^{N_c}\sum_{p=1}^{P} -2\mathbf{Re}\left[\by^* \bPo \nabla_{{\varphi}_{c,p}}\bH \bG^{-1}\bH^*\by \right]d\varphi_{c,p} \label{eqn:dJ'_3}
\end{align}
We then apply a similar first order expression for the differential $dJ = \sum_{c = 1}^{N_c}\sum_{p=1}^{P} \nabla_{{\varphi}_{c,p}} J d\varphi_{c,p}$. By comparing this expansion with Equation \eqref{eqn:dJ'_3}, we see that the gradient of $J$ with respect to each individual parameter $\varphi_{c,p}$ is
\begin{align} \label{eqn:delJ_final}
\nabla_{{\varphi}_{c,p}} J = -2\mathbf{Re}\left[\by^* \bPo \nabla_{{\varphi}_{c,p}}\bH \bG^{-1}\bH^*\by\right] 
\end{align}

\noindent Here, $\nabla_{{\varphi}_{c,p}}\bH$ contains the element-wise partial derivatives of $\bH$ with respect to each parameter. Recall that $\bH = \bH(\bphi) = \left[\bH(\bphi_1) \, \dots \, \bH(\bphi_{N_c})\right]$. Therefore, 
\begin{align}
    \delphicp\bH(\bphi) &= \delphicp \left[\bH(\bphi_1) \, \dots \, \bH(\bphi_{N_c})\right] \nonumber \\
    &= \left[\delphicp\bH(\bphi_1) \, \dots \, \delphicp\bH(\bphi_{N_c})\right] \nonumber \\
    &= \left[\boldsymbol{0} \, \dots \delphicp \bH(\bphi_c) \dots \, \boldsymbol{0} \right] \nonumber \\
    &= \left[\boldsymbol{0} \, \dots \, \left[\delphicp\bh_{c,0} \, \dots \, \delphicp\bh_{c,A_c}\right]  \dots \boldsymbol{0} \right] \nonumber \\
    &= \left[\boldsymbol{0} \, \dots \, j 2\pi \bn \odot \left[\bh_{c,0} \, \dots \, \bh_{c,A_c}\right]  \dots \boldsymbol{0} \right] \label{eqn:delH_final}
\end{align}
where $\bn = \left[\left(\tfrac{0}{f_s}\right)^p \, \dots \, \left(\tfrac{N-1}{f_s}\right)^p \right]^T$ and $\odot$ denotes element-wise multiplication. The complete expression for the gradient $\nabla_{{\varphi}_{c,p}} J$ can be obtained by substituting Equation \eqref{eqn:delH_final} into Equation \eqref{eqn:delJ_final}.

\end{document}